\documentclass[preprint]{elsarticle}

\usepackage{hyperref}
\usepackage{amsmath,amssymb}
\usepackage{siunitx}
\usepackage{graphicx}
\usepackage{upgreek}

\usepackage{geometry}
\journal{Solid-State Electronics}

\begin{document}

\begin{frontmatter}

\title{Characterization and Modeling of 28-nm FDSOI CMOS Technology down to Cryogenic Temperatures}


\author{$^\dagger$Arnout Beckers}
\cortext[mycorrespondingauthor]{Corresponding author}
\ead{arnout.beckers@epfl.ch}
\author{$^\dagger$Farzan Jazaeri}
\author{$^\ddagger$Heorhii Bohuslavskyi}
\author{$^\ddagger$Louis\,Hutin}
\author{$^\ddagger$Silvano\,De\,Franceschi}
\author{$^\dagger$Christian Enz}
\address{$^\dagger$Integrated Circuits Laboratory (ICLAB), Ecole Polytechnique F\'ed\'erale de Lausanne (EPFL), Switzerland}
\address{$^\ddagger$CEA-L\'eti, Grenoble, France}

\tnotetext[t1]{This project has received funding from the European Union's Horizon 2020 	Research $\&$ Innovation Programme under grant agreement No. 688539 MOS-Quito. }
\vspace{-10cm}

\begin{abstract}
This paper presents an extensive characterization and modeling of a commercial 28-nm FDSOI CMOS process operating down to cryogenic temperatures. The important cryogenic phenomena influencing this technology are discussed. The low-temperature transfer characteristics including body-biasing are modeled over a wide temperature range (room temperature down to 4.2\,K) using the design-oriented simplified-EKV model. The trends of the free-carrier mobilities versus temperature in long and short-narrow devices are extracted from dc measurements down to 1.4\,K and 4.2\,K respectively, using a recently-proposed method based on the output conductance. A cryogenic-temperature-induced mobility degradation is observed on long $p$MOS, leading to a maximum hole mobility around 77\,K. This work sets the stage for preparing industrial design kits with physics-based cryogenic compact models, a prerequisite for the successful co-integration of FDSOI CMOS circuits with silicon qubits operating at deep-cryogenic temperatures. 	
\end{abstract}


\end{frontmatter}

\section{Introduction}
The birth of CMOS-compatible qubits in silicon~\cite{pla2012single,elzerman2004single,veldhorst2014addressable,maurand_cmos_2016} has rebooted the interest in cryogenic CMOS electronics for computing applications. Since the 1970s, 
MOSFET devices have been under investigation at cryogenic temperatures for use in custom applications, such as low-noise scientific equipment, spacecraft, power conversion etc.~\cite{kirschman1985cold,gutierrez2000low,balestra_device_2001,claeys_perspectives_1994,elewa}. However, despite its many benefits for reaching high-performance and low-power computing~\cite{deen}, cryogenic cooling did not stay into practice for computing, abandoning the trend set by the ETA-10 liquid-nitrogen-cooled supercomputer~\cite{eta}. 

Nowadays, co-integrating qubits and CMOS circuits on the same substrate  can greatly aid the development of scalable quantum computers featuring massive parallelism and error correction\cite{ekanayake_characterization_2010,reilly2015engineering,vandersypen2017interfacing}. In this context, a silicon-on-insulator (SOI) platform is particularly attractive since the back gate provides additional control over the electron-spin qubit, trapped under the front gate of a SOI (nanowire) MOSFET\cite{ekanayake_characterization_2010,hutin_soi_2017,franceschi_soi_2016}. To integrate the control circuits with quantum devices working at deep-cryogenic temperatures, regular SOI MOSFETs need to demonstrate reliable digital, analog and RF functionalities at such low temperatures. Using SOI cryogenic control electronics, the back gate can prove a useful tool to control the threshold voltage and hence the power consumption in circuits integrated close to the qubits~\cite{bohus}, benefiting qubit coherence time by lowering generated noise. The main focus is on advanced ultra-thin body fully-depleted SOI (FDSOI) technology, e.g., the 28-nm node, to enable ultimate scalability of the resulting hybrid quantum-classical system~\cite{franceschi_soi_2016,clapera}. 

The 28-nm node, presently considered the ideal node for analog and RF applications at room temperature~\cite{cathelin}, has recently been tested for digital and analog functionality down to liquid-helium temperature (4.2\,K)~\cite{bohus}, and millikelvin temperature (20\,mK) \cite{galy}. The improvement in RF characteristics has  been verified down to liquid-nitrogen temperature (77\,K)~\cite{kazemi}. In addition to device characterization, it is mandatory that industry-standard compact models~\cite{utsoipartone,utsoiparttwo,bsimimg} become compatible with cryogenic temperatures, to achieve optimal cryogenic CMOS designs controlling a large number of qubits. To date, important temperature-related phenomena have been included only by fitting the characteristics using the existing temperature-scaling laws available in industry-standard compact MOS transistor models dedicated to room-temperature operation, i.e., for bulk~\cite{akturk} and double-gate MOSFET~\cite{akturk2}.  However, this approach cannot provide a physically-sound basis to further develop compact models targeting reliable CMOS designs at cryogenic temperatures. Recently, an analytical model for bulk cryogenic MOSFET operation~\cite{tedpaper} has been proposed, which has been developed starting from the Poisson equation at cryogenic temperatures, validating the Boltzmann statistics and taking into account the temperature dependencies of dopant freeze-out, bandgap widening and the Fermi-Dirac occupation of interface charge traps. This model provides the necessary analytical physics-based expressions for compact modeling purposes at cryogenic temperatures. Furthermore, a body-partitioning technique has been developed which can be used to calculate the extension of the ionized layer of dopants under the gate by field-assisted ionization when the substrate is initially frozen-out at cryogenic temperatures~\cite{beckers_jeds}. From this method, it can be expected that the low-doped thin film of silicon in a FDSOI MOSFET can be completely ionized at cryogenic temperature depending on the relative bias points of the front and back gates.  

In this work, as an initial investigation prior to further physical and compact modeling, we perform a cryogenic characterization and semi-empirical modeling of a commercial ultra-thin-body 28-nm FDSOI CMOS technology at temperatures down to 4.2\,K, similar to earlier work on a commercial 28-nm bulk CMOS technology~\cite{essderc}. The low-temperature dc measurements (transfer and output characteristics) and the characterization down to 4.2\,K are presented in Sec.\,\ref{sec:meas} and \ref{sec:char}, respectively. In Sec.\,\ref{sec:phenomena}, we qualitatively explain the influence of cryogenic temperatures on the electrical behavior of this technology, with a main focus on incomplete ionization (freeze-out) and interface charge traps. The free-carrier mobility trends versus temperature are obtained from the recently-proposed $g_{ds}$-function method~\cite{jazaeri}, which is derived here for FDSOI technologies (Sec.\,\ref{sec:mob}). This method allows to extract the mobility from dc transfer or output measurements. In Sec.\,\ref{sec:modeling}, we illustrate how the drastic temperature reduction, the incomplete ionization, and the interface charge traps can be accounted for in circuit device-models, taking as an example the simplified-EKV model, a simple design-oriented model using only four parameters for short-channel devices.  

\section{\label{sec:meas}Low-Temperature Measurements} 
\begin{figure}[t]
	\includegraphics[width=\textwidth]{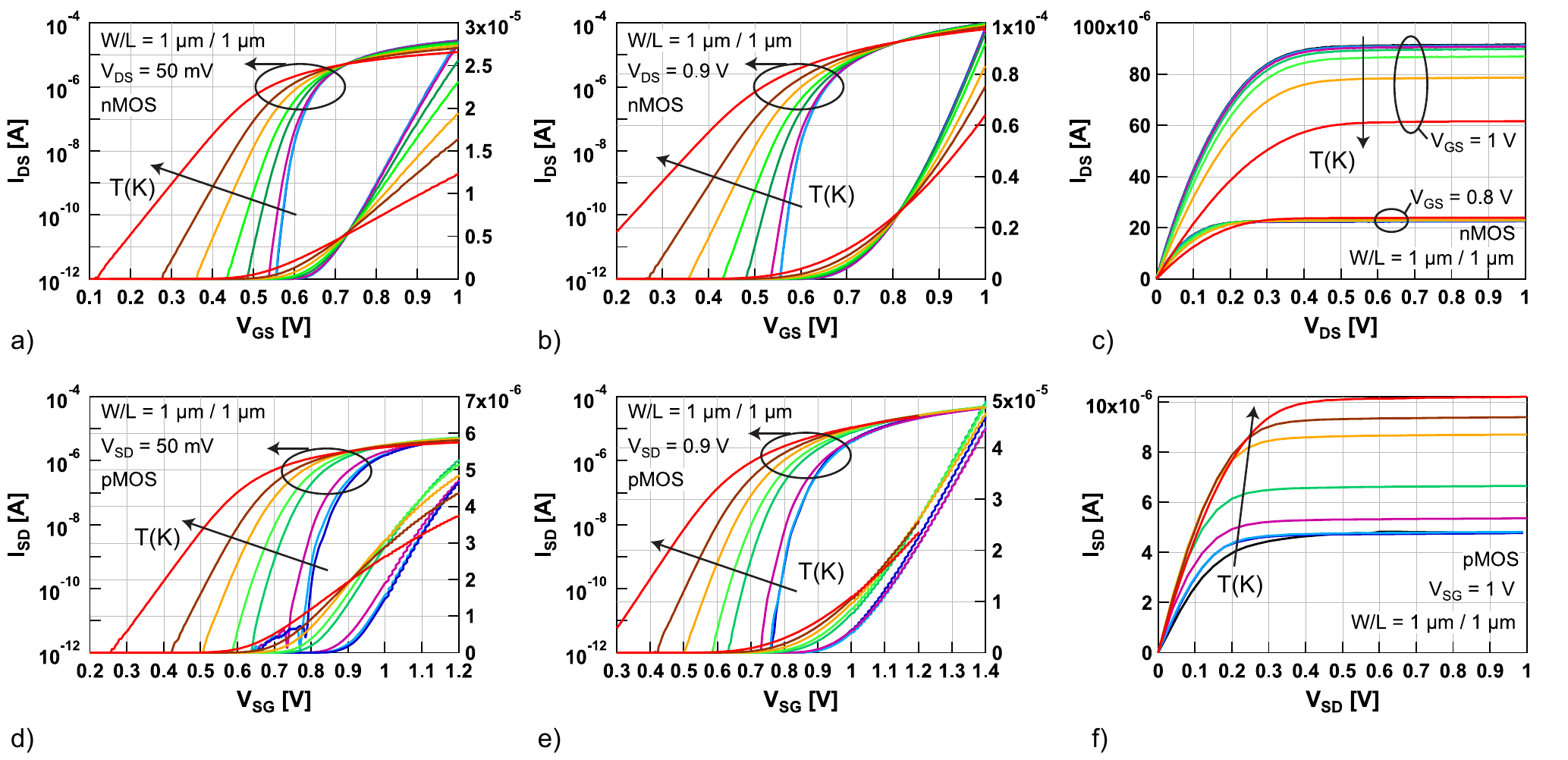}
	\vspace{-0.7cm}
	\caption{Transfer and output characteristics measured in wide-long $n$MOS and $p$MOS devices ($W/L=1\mu m/1\mu m$) of a 28-nm FDSOI CMOS technology at zero back-gate voltage. The linear and saturation transfer characteristics are presented down to 4.2\,K (red: 300\,K, brown: 210\,K, orange: 160\,K, light green: 110\,K, dark green: 77\,K, purple: 36\,K, light blue: 10\,K, and dark blue: 4.2\,K). In subfigures a, b, and e the curves at 10\,K and 4.2\,K lie almost on top of each other. The output characteristics (c and f) are presented down to 1.4\,K (black: 1.4\,K). At $V_{\mathrm{GS}}=1$\,V, the saturation current decreases with decreasing temperature for $p$MOS, while it increases for $n$MOS. This hints on a cryogenic-temperature-induced mobility degradation in long $p$MOS devices, as will be demonstrated in Sec.\,\ref{sec:mob}. The temperature dependencies of the free-carrier mobilities can be extracted from c) and f) knowing that the mobility is proportional to the derivative of the output conductance at small $V_{\mathrm{DS}}$ (see Sec.\,\ref{sec:mob}).}
	\label{figure1}
\end{figure}

Transfer characteristics in linear ($\vert V_{\mathrm{DS}}\vert$\,=\,50\,mV) and saturation ($\vert V_{\mathrm{DS}}\vert$\,=\,0.9\,V) were measured on various devices of a 28-nm FDSOI CMOS process from room temperature down to 4.2\,K, including changes in the body bias~\cite{bohus}. Intermediate temperature steps were taken at 10, 36, 77, 110, 160, and 210\,K, and the back-gate voltage ($V_{back}$) was ramped from $-$0.9\,V to 0.9\,V. Output characteristics were measured at zero $V_{back}$ using the same temperature steps, and additionally at 1.4\,K for some devices. Figures (\ref{figure1}), (\ref{figure2}), and (\ref{figure3}) show these measurements for large (Fig.\,\ref{figure1}), narrow (Fig.\,\ref{figure2}), and small (Fig.\,\ref{figure3}) $n$MOS and $p$MOS devices. 

\subsection{Discussion}
A clear improvement in the subthreshold swing and transconductance is evident for the wide-long $n$MOS and $p$MOS devices in Fig.\,\ref{figure1}. However, it should be noted that the improvement is minimal between 10\,K and 4.2\,K with curves lying almost on top of each other. The on-state current increases with decreasing temperature in long $n$MOS, but decreases in long $p$MOS, as highlighted by the opposite temperature trends in the output characteristics in Figs.\ref{figure1}-c and \ref{figure1}-f. This will be explained by a cryogenic-temperature-induced mobility degradation in $p$MOS in Sec.\,\ref{sec:mob}. The fact that the initial slope of the output characteristics in the linear regime (small $V_{\mathrm{DS}}$) changes with temperature can be used to extract the mobility trend versus temperature down to 1.4\,K, according to the $g_{ds}$-function method described in Sec.\ref{sec:mob}.  Figure \ref{figure2} shows similar temperature-dependent dc characteristics for narrow-short and narrow-long devices. However, conductance oscillations are observed on a narrow-short $p$MOS (Fig.\,\ref{figure2}-d) in the deep-cryogenic range starting from 36\,K. These oscillations have been attributed to the presence of dopants diffused from source and drain into the channel~\cite{bohus}. It can be noted that the oscillations becomes less pronounced with increasing temperature, gradually disappearing at 77\,K and 110\,K (green curves). In Fig.\,\ref{figure2} (linear scale) at high gate voltages an impact of access resistance or mobility degradation due to the vertical field is noticeable on narrow-short $n$MOS in the linear regime. In Fig.\ref{figure3}, for wide-short $p$MOS, conductance oscillations are also observed. In the output characteristics (Fig.\,\ref{figure3}-c and \ref{figure3}-f) a drain-induced-barrier-lowering is present at all temperatures, which is roughly temperature dependent. No kink effect is observed in this advanced fully-depleted technology. Since the output conductance in saturation is almost constant with temperature, the intrinsic gain versus temperature will follow the increase in transconductance with decreasing temperature.

\subsection{\label{sec:phenomena}Low-Temperature Phenomena} 
Important cryogenic phenomena influencing double-gate MOSFET performance have been extensively reviewed by Balestra and Ghibaudo \cite{briefreview,balestra2017physics,balestra_device_2001}, and Claeys and Simoen\cite{claeys_perspectives_1994}. These phenomena, also present at room temperature but to a lower degree, include interface traps, dopant incomplete ionization, field-assisted ionization, mobility temperature-trend, bandgap temperature-trend, exponential temperature dependency of the intrinsic carrier concentration, and quantum effects. It should be noted that the kink effect in the output characteristics, prominently present in older technologies at cryogenic temperatures~\cite{kink}, has not been observed in this fully-depleted technology below the used supply voltage. Below follows a brief description of the phenomena which can impact a fully-depleted FDSOI technology, and how to model them: 
\begin{figure}[t]
	\includegraphics[width=\textwidth]{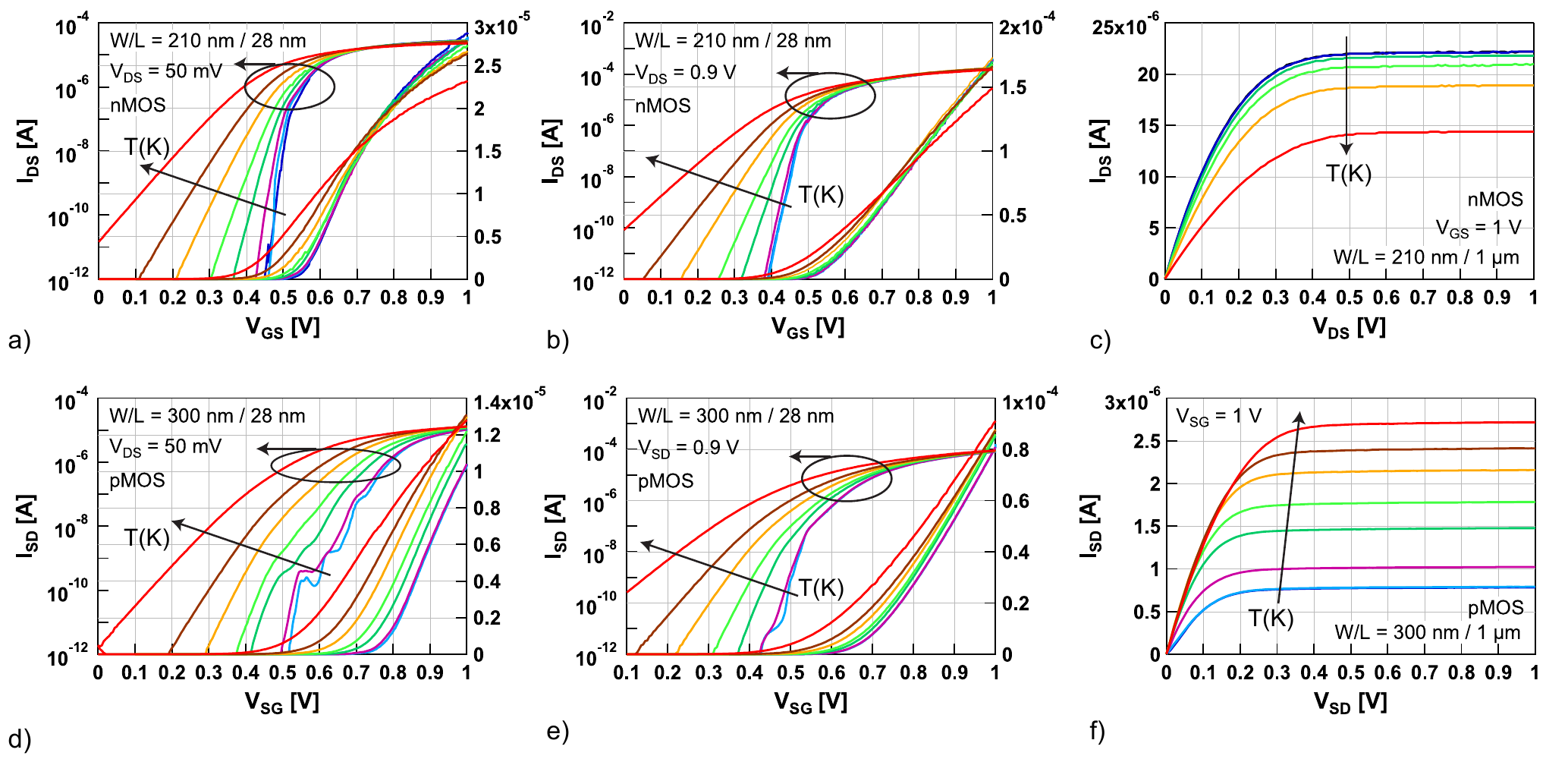}
	\vspace{-0.7cm}
	\caption{a)-b) Transfer characteristics measured in a narrow-short ($W/L=$210\,nm/28\,nm) $n$MOS device down to 4.2\,K at zero back-gate voltage, c) Output characteristics measured in a narrow-long ($W/L=$210\,nm/1\,$\mu$m) $n$MOS device down to 1.4\,K, d)-e) Transfer characteristics measured on narrow-short $p$MOS ($W/L=$300\,nm/28\,nm) down to 4.2\,K at zero back-gate voltage, f) Output characteristics measured in a narrow-long $p$MOS ($W/L=$300\,nm/1\,$\mu$m) down to 1.4\,K, The color scheme for the intermediate temperatures is the same as in Fig.\,\ref{figure1}.}
	\label{figure2}
\end{figure}
\begin{figure}[t]
	\includegraphics[width=\textwidth]{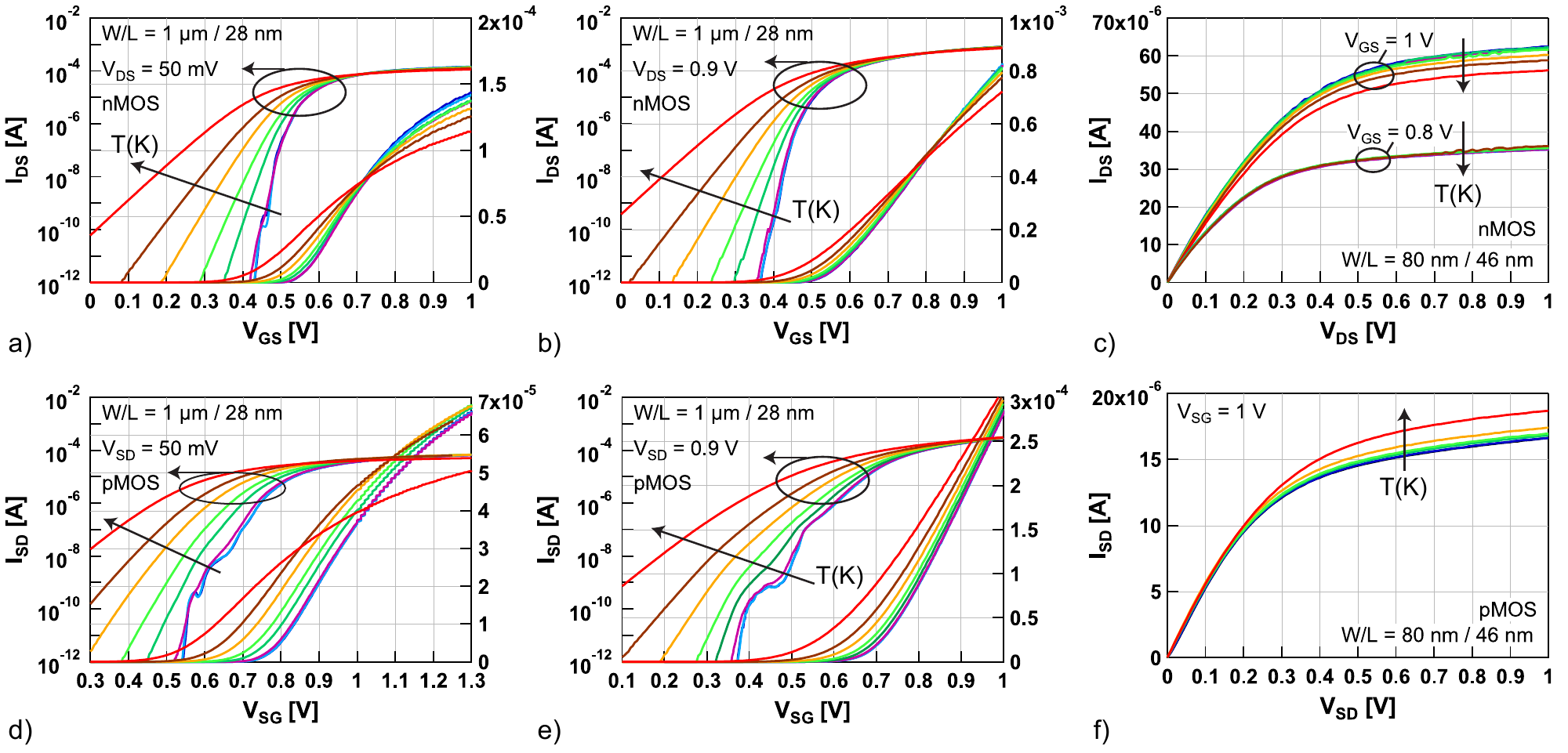}
	\vspace{-0.7cm}
	\caption{a)-b) Transfer characteristics measured in a wide-short ($W/L=$1\,$\mu$m/28\,nm) $n$MOS device down to 4.2\,K at zero back-gate voltage, c) Output characteristics measured in a small $n$MOS device with $W/L=$80\,nm/46\,nm down to 1.4\,K at two different gate voltages, d)-e) Transfer characteristics measured in a wide-short $p$MOS ($W/L=$1\,$\mu$m/28\,nm) down to 4.2\,K at zero back-gate voltage, f) Output characteristics measured in a small $p$MOS ($W/L=$80\,nm/46\,nm) down to 1.4\,K at zero back-gate voltage. The color scheme for the intermediate temperatures is the same as in Fig.\,\ref{figure1}.}
	\label{figure3}
\end{figure}
\begin{itemize}
	\item \textit{Incomplete ionization} or \textit{substrate freeze-out}
	In a MOSFET in thermal equilibrium (no voltage applied) dopant atoms will become deionized at a sufficiently low (cryogenic) temperature depending on the doping concentration in the range $\SI{e12}{}$ to $\SI{e18}{\per\centi\meter\cubed}$ ~\cite{foty_impurity_1990,simoen1989freeze, pierret1987advanced,substrate}. An overview of the freeze-out critical temperatures for each doping concentration in this range in silicon can be found in~\cite{beckers_jeds}. At higher doping concentrations, e.g., in the source and drain contacts, no freeze-out happens due to the formation of impurity bands which overlap with the conduction or valence band edges~\cite{jonsher}. Therefore, freeze-out does not influence the access resistance improvement at cryogenic temperatures~\cite{	akturkeffects}. For a $p$-type silicon body, an acceptor dopant atom will be ionized from a theoretical viewpoint when the acceptor energy $E_A$ is occupied by an electron. Therefore, the ionized dopant concentration, $N_A^-$, is given by
	\begin{equation}\label{inco}
	N_A^-=N_Af(E_A)=\frac{N_A}{1+g_Ae^{\frac{E_A-E_{F,n}}{kT}}}=\frac{N_A}{1+g_Ae^{\frac{\psi_A-(\psi-V_{ch})}{U_T}}}, 
	\end{equation}
	where $f(E_A)$ is the ionization probability given by a Fermi-Dirac distribution function, and the electron quasi-Fermi-level is $E_{F,n}=E_F-qV_{ch}$ with $V_{ch}$ the channel voltage. The RHS of (\ref{inco}) is convenient for direct inclusion in the Poisson-Boltzmann equation~\cite{tedpaper}.
	
	In FDSOI, the doping concentration is rather low ($\approx \SI{e15}{\per\centi\meter\cubed}$) compared to the inversion charge density. However, as illustrated in Fig.\,\ref{figure4}-a, in the flatband condition and at 4.2\,K, approximately all dopants will be frozen-out, independent of the doping concentration in the range ($\SI{e12}{}-\SI{e18}{\per\centi\meter\cubed}$). The calculated $E_F$-position at 4.2\,K lies under $E_A$, leading to freeze-out or $f(E_A)\ll1$ (Fig.\,\ref{figure4}-b). Nonetheless, the front-gate voltage will ionize the impurities under the surface of the front-gate, when $E_A$ bends under $E_F$ near the surface of the front-gate. In the subthreshold region, when $E_F\approx E_c-3U_T$, complete ionization can be assumed under the front gate. This transition from freeze-out to complete ionization	due to the applied field can lead to a kink in early depletion~\cite{tedpaper}. Note that depending on the band bending at the front and back gates in a certain mode of operation, it is possible that the dopants under the front gate are completely ionized but frozen-out under the back gate  or vice versa. A complete comprehension of the field-assisted ionization effect on the mobile charge density in FDSOI below inversion and including body bias, would require a more in-depth physical analysis. 
	
	\begin{figure}[h]
		\centering
		\includegraphics[width=0.9\textwidth]{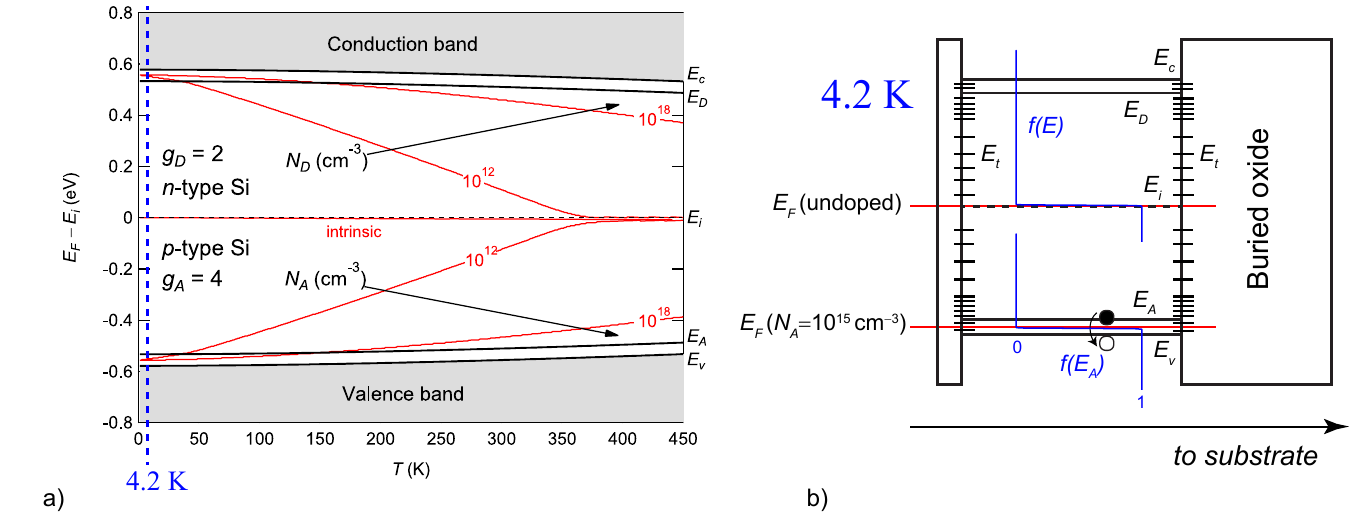}
		\vspace{-0.3cm}
		\caption{a) Simulated position of the Fermi-level (red) in $n$- and $p$-type doped silicon as a function of temperature and doping concentration. At 4.2\,K, the silicon is frozen-out for all doping concentrations in the range (\SI{e12}{}\,-\,\SI{e18}{\per\centi\meter\cubed}) since $E_F>E_D$ or $E_F<E_A$, and $f(E_D)$ and $f(E_A)$ are close to step functions. The temperature dependency of the bandgap from Varshni~\cite{varshni} is used, b) Illustration of freeze-out and interface charge traps in the thin silicon film of a FDSOI MOSFET with a $p$-type body ($N_A=\SI{e15}{\per\centi\meter\cubed}$). The two phenomena can be described by Fermi-Dirac statistics. The position of the Fermi level is shown in red. The probabilities of dopant ionization and interface-trap occupation depend on the position of the Fermi-level in the silicon film with respect to $E_A$ and $E_{t}$, respectively. In case $E_A$ bends under $E_F$ near the surface of one of the gates, an ionized layer of dopants forms under the gate. Field-assisted ionization makes $f(E_A)\approx 1$ before inversion is reached. In the figure it is assumed that the front and back gates are biased such that a flatband situation is created. }
		\label{figure4}
	\end{figure}	

	\item \textit{Temperature-dependent occupation of interface charge traps}\cite{hafez_assessment_1990,trevisoli_junctionless_2016}
	Interface traps need to be included for both the front and the back gate, as illustrated in Fig.\,\ref{figure4}-b. This adds two additional Fermi-Dirac temperature dependencies $f(E_{t,j})$ (with $E_{t,j}$ a trap energy-level at position $j$ in the bandgap), apart from the ionization probability $f(E_A)$. The interface traps can be modeled as a discrete summation of traps, as explained in \cite{yesayan2016charge,jazaeri_sallese_2018,jazaeri22}. The temperature-dependent occupation of interface traps is important for a correct derivation of the subthreshold-swing formula, leading to hyperbolic temperature dependency of the slope factor (ignoring coupling effects between front and back gates), which will be discussed in more detail in Sec.\,\ref{sec:char}. 
	
	\item \textit{Bandgap widening} The total change of the silicon bandgap from room temperature down to 4.2\,K is approximately 1.12 to 1.16 eV, widening with decreasing temperatures~\cite{varshni}. The temperature dependence in the cryogenic regime ($<100$\,K) is almost constant, as illustrated in Fig.\,\ref{figure4}-a.  
	\item \textit{Exponential temperature dependency of the intrinsic carrier concentration} The intrinsic carrier concentration is given by $n_i=\sqrt{N_cN_v}\exp\left[-E_g/(2kT)\right]$, which at 4.2\,K leads to extremely small values lying outside IEEE double-precision arithmetic, in the order of $\SI{e-650}{\per\centi\meter\cubed}$~\cite{tedpaper}. This is physically accurate since the overlap of a Fermi-Dirac function at 4.2\,K, lying at the intrinsic level (Fig.\,\ref{figure4}-b), and the density-of-states in the conduction band becomes very small. 

	\item \textit{Quantum confinement and quantum transport} Quantum effects become more pronounced in FDSOI MOSFETs at cryogenic temperatures, since they are less obscured by thermal fluctuations when the quantized energy is similar to the thermal energy~\cite{colinge_low-temperature_2006}. 
\end{itemize}

\begin{figure}[t]
	\centering
	\includegraphics[width=\textwidth]{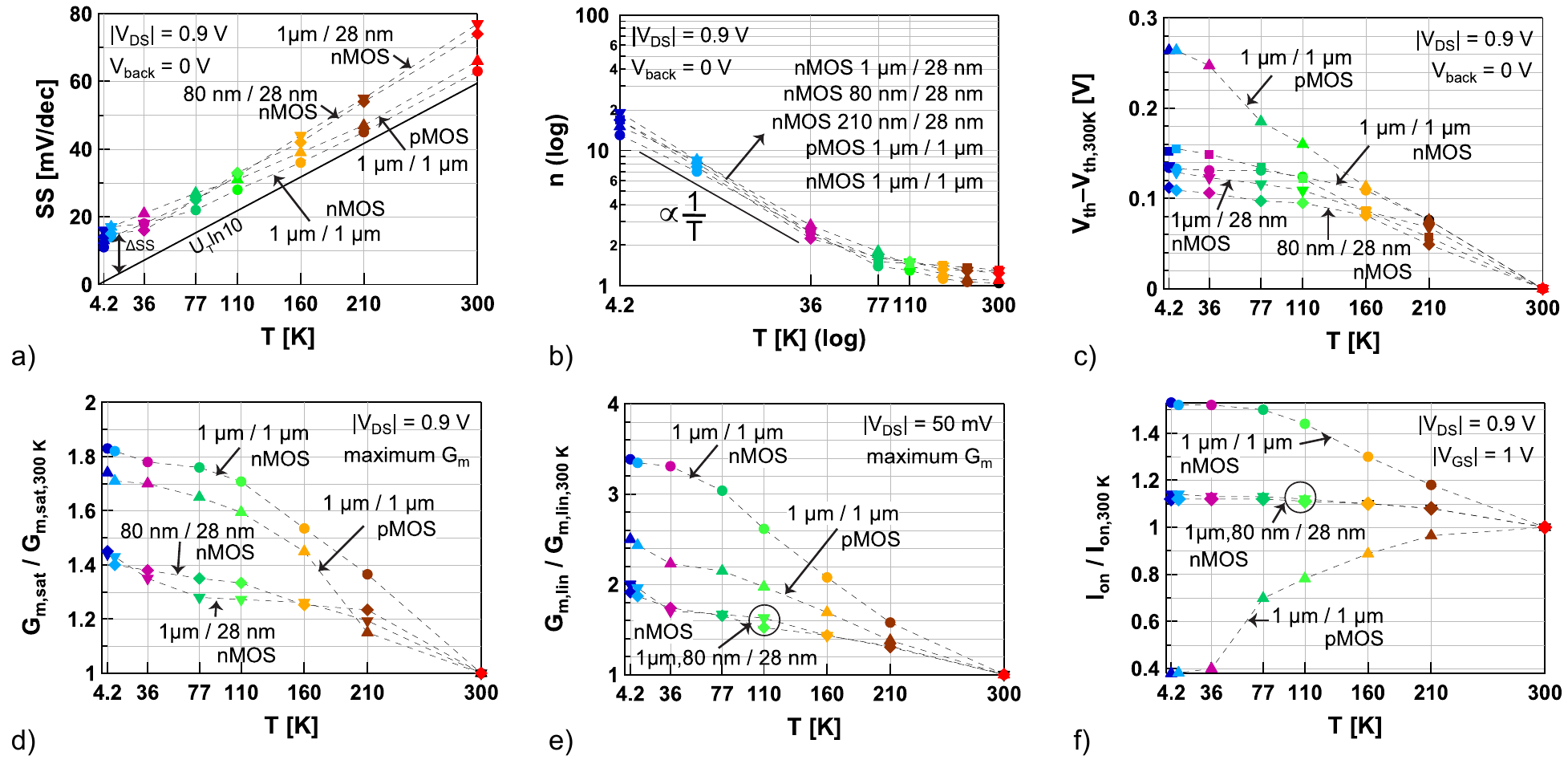}
	\caption{Characterization of a 28-nm FDSOI CMOS technology down to liquid-helium temperature (4.2\,K), a) Subthreshold swing $SS$, b) Slope factor ($n=SS/(U_T\ln10)$, c) Threshold voltage shift with respect to room temperature $\Delta V_{th}=V_{th}-V_{th,300}$, d) Transconductance in saturation, e) Transconductance in linear, f) Ratio of low-temperature over room-temperature on-state current.}
	\label{fig:char}
\end{figure}

\section{\label{sec:char}Characterization}
In this section, the following technological parameters are extracted from the cryogenic measurements: the subthreshold swing ($SS$), slope factor ($n$), threshold voltage ($V_{th}$), transconductance in linear and saturation ($G_{m,lin},G_{m,sat} $), the on-state current ($I_{on}$), and the effective free-carrier (electron and hole) mobilities ($\mu_{eff}$).

\subsection{Subthreshold swing, Threshold voltage, Transconductance, and On-state current}

As illustrated in Fig.\,\ref{fig:char}-a, for temperatures below $\approx$\,160\,K, the extracted average $SS$-values show an increasing offset, $\Delta SS$, from the thermal limit, $U_T\ln10$, with $U_T\triangleq kT/q$ the thermal voltage. $\Delta SS$ reaches around 10\,mV/dec at 4.2\,K for long $n$MOS, since $U_T\ln10$ predicts $\approx$\,0.8\,mV/dec. The slope factors required to reach such high $SS$-values are shown in Fig.\,\ref{fig:char}-b ($n=SS/(U_T\ln10)$). From this figure a hyperbolic temperature-dependency of $n$ is evident, which is not strongly dependent on geometry at cryogenic temperatures. The values below 77\,K in Fig.\,\ref{fig:char}-a cannot be explained anymore by $n_0U_T\ln10$, where the slope-factor $n_0$ is limited by two, according to $n_0=1+C_{dep}/C_{ox}$. Here the depletion capacitance $C_{dep}$ is smaller than the oxide capacitance $C_{ox}$. Furthermore, including the interface-trap capacitance, $C_{it}=qN_{it}$, i.e. $n_0=1+(C_{dep}+C_{it})/C_{ox}$ with $N_{it}$ the density-of-interface-traps per unit area, would lead to very high extracted values for $N_{it}$ in the order of $\SI{e13}{\per\square\centi\meter}$ at 4.2\,K~\cite{elewa,trevisoli_junctionless_2016,hafez_assessment_1990}, and $\SI{e17}{\per\square\centi\meter}$ at 20\,mK~\cite{galy}. The latter is higher than the density of surface-states in silicon ($\SI{e15}{\per\square\centi\meter}$). However, it should be emphasized that in this $n_0$-formula the temperature-dependent occupation of interface-traps is not taken into account (see Sec.\,\ref{sec:phenomena}). Reliable extraction of the interface-trap-density at deep-cryogenic and millikelvin temperatures using the standard $SS$-formula is therefore questionable. Inclusion of the interface-trap temperature dependency into the subthreshold swing theory of bulk MOSFET has been shown to yield lower extracted $N_{it}$-values~\cite{tedpaper}. Similarly to the derivation for bulk MOSFET presented in \cite{tedpaper}, by including $f(E_t)$ the temperature dependency of $n \propto 1/U_T$ can be derived for the front-gate in FDSOI as well, ignoring the coupling effects between front and back gates. This gives $SS=n(T)U_T\ln 10=n_0U_T\ln10+\Delta SS$, where $n_0$ is the slope factor without interface traps, and $\Delta SS$ the subthreshold-swing offset as observed in Fig.\,\ref{fig:char}-a. $\Delta SS$ is given by $(qN_{it}/C_{ox})\ln10\left[g_t/(1+g_t)^2\right]$ with $N_{it}$ the density-of-interface-traps and $g_t$ the trap degeneracy factor. Note that in this model, $N_{it}$ does not become multiplied with $U_T$, resulting in reasonable extracted values for $N_{it}$ at cryogenic temperatures lower than found in  \cite{hafez_assessment_1990,trevisoli_junctionless_2016,galy}. The $\Delta SS$-offset starts to increase below $\approx$\,160\,K since the subthreshold region happens when $E_F$ lies closer to $E_c$, where $N_{it}$ is observed to be higher already at 300\,K (see also Fig.\,\ref{figure4}-b)~\cite{tewksbury}.

The shift in threshold voltage at 4.2\,K with respect to room temperature increases in the order of 0.1$-$0.3\,V (Fig.\,\ref{fig:char}-c). Note that the largest $V_{th}$-increase is observed for $p$MOS, similarly to a 28-nm bulk process\cite{essderc}. Furthermore, the maximum $G_{m,sat}$ and $G_{m,lin}$ (Figs.\ref{fig:char}d-e) improve down to 4.2\,K, e.g.\,respectively $\times$\,3.4 (linear) and $\times$\,1.8 (saturation) for $n$MOS $W/L$\,=\,1 $\upmu$m\,/\,1\,$\upmu$m. In Fig.\ref{fig:char}-f, $I_{on}$ is extracted at $\vert V_{GS}\vert=$\,1\,V. Note that the actual trend of $I_{on}$ with temperature is strongly dependent on the bias and the device-type. At a standard supply voltage of 1\,V, the on-state current increases with decreasing temperature for long $n$MOS (Fig.\,\ref{figure1}-a-c), while it decreases for $p$MOS (Fig.\,\ref{figure1}-d-f). However, a cryogenic-temperature-induced mobility degradation has not been extracted from measured CV characteristics on this device~\cite{bohus}. Therefore, in the next section, we take a second look at the characterization of the free-carrier mobility in this technology at cryogenic temperatures. 

\subsection{\label{sec:mob}Free-Carrier Mobility}
In doped bulk silicon, the free-carrier mobility is expected to drop when transitioning below a certain cryogenic temperature and Coulombic impurity scattering becomes dominant over phonon scattering, leading to a typical bell-shaped mobility trend with respect to temperature\cite{jonsher,kirschman1985cold} This behavior can be different in MOSFET devices when the channel is ballistic.

The free-carrier mobility in MOSFETs is usually extracted from dc measurements using the $Y$-function approach~\cite{ghibaudo, emrani,shin_low_2014}, or from a combination of dc and capacitance measurements using the split-$\mathrm{CV}$ method. For advanced CMOS technologies, the split-CV method can yield unreliable results due to the dominance of extrinsic capacitances, and the $Y$-function has two effective slopes. Both methods are also used to characterize advanced devices down to deep-cryogenic ($<$ 10\,K) temperatures \cite{huang,gildy,balestra2017physics,bohus}. Dopant freeze-out and field-assisted ionization may change the dependency of the scattering mechanisms on the gate voltage. Since in strong inversion field-assisted ionization is complete, the underlying assumption of the $Y$-function approach, i.e., the homographic gate-voltage dependent mobility law, $\mu=\mu_0/\left[1+\theta(V_{\mathrm{GS}}-V_t)\right]$\cite{ghibaudo}, can still provide an adequate description of the mobility down to cryogenic temperatures. On the other hand, the split-$\mathrm{CV}$ method  has to deal with the unknown thermal behavior of the extrinsic capacitances, and requires deep-cryogenic cooling for two types of measurements. Using CV measurements, a mobility degradation at cryogenic temperatures has not been observed for long $p$MOS in this technology~\cite{bohus}. In the next subsection, we obtain the free-carrier mobilities from dc measurements according to an approach recently developed by Jazaeri et al\cite{jazaeri}. 

\begin{figure}[t]
	\centering
	\includegraphics[width=\textwidth]{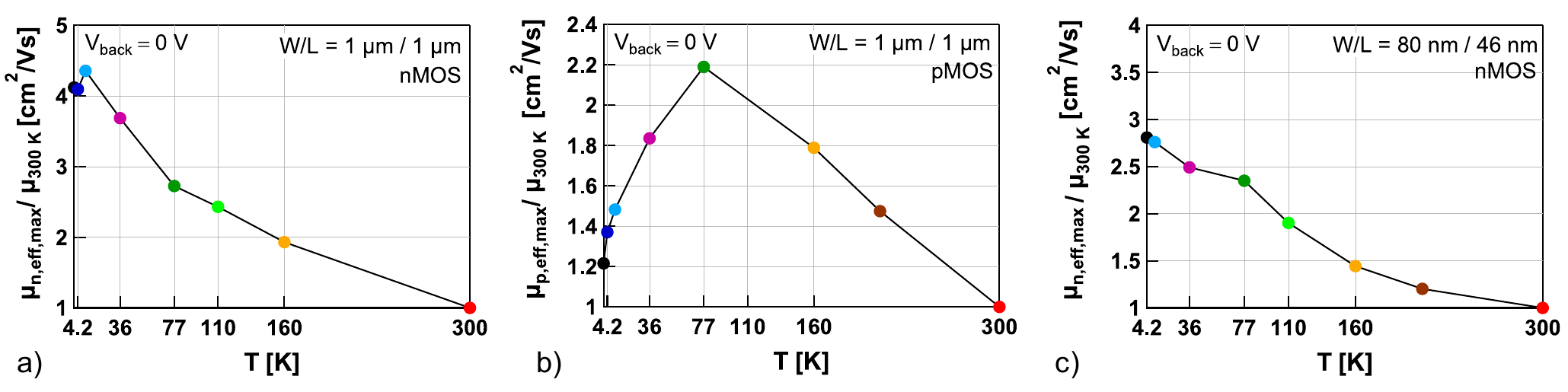}
	\vspace{-0.7cm}
	\caption{Free-carrier mobilities in a 28-nm FDSOI CMOS process extracted using the $g_{ds}$-function~\cite{jazaeri} and normalized to the room-temperature mobility, a) $n$MOS 1\,$\mu$m/1\,$\mu$m down to 1.4\,K, b) $p$MOS 1\,$\mu$m/1\,$\mu$m down to 1.4\,K, and c) $n$MOS 1\,$\mu$m/28\,nm down to 4.2\,K. A free-carrier mobility degradation induced by cryogenic temperatures is extracted on long $p$MOS starting from around 77\,K, and around 10\,K in long $n$MOS. For the considered 28-nm process~\cite{bohus}, the EOT is 1.55\,nm for $n$MOS, and 1.7\,nm for $p$MOS with thin front-gate oxide~\cite{bohus}.}
	\label{figure:mob}
\end{figure}

\subsection{Free-Carrier Mobility Extraction using $g_{ds}$-function}

The Jazaeri mobility-extraction method ($g_{ds}$-function)~\cite{jazaeri} does not assume any gate-voltage dependent mobility law \textit{a priori}. The method only assumes drift-diffusion transport in advanced field-effect transistors as a starting point of the derivation~\cite{jazaeri}. Drift-diffusion transport has been shown to give an accurate representation of the current down to 4.2\,K, and the validity of the Boltzmann statistics has been demonstrated down to millikelvin temperatures~\cite{beckers_jeds,tedpaper}. Therefore, here we can extend this method to deep-cryogenic temperature operation. In what follows, the method will be briefly derived for SOI technology. 

Following the approach in \cite{jazaeri}, for two different operating points in the linear regime, drift-diffusion gives $I_{D1}=-(W/L_G)\mu\overline{Q}_{m1}V_{\mathrm{DS1}}$, and $I_{D2}=-(W/L_G)\mu\overline{Q}_{m2}V_{\mathrm{DS2}}$, 
where $\overline{Q}_m$ is the mean value of the local mobile charge densities at source and drain, i.e., $\overline{Q}_m\triangleq (Q_{mS}+Q_{mD})/2$, and it is assumed that $\mu$ is not as a function of $V_{\mathrm{DS}}$ for small $V_{\mathrm{DS}}$. Hence we can derive that 
\begin{equation}
\frac{1}{\overline{Q}_{m2}}\frac{\partial \overline{Q}_m}{\partial V_{\mathrm{DS}}}\approx \left(\frac{I_{\mathrm{D1}}-I_{\mathrm{D2}}}{V_{\mathrm{DS1}}-V_{\mathrm{DS2}}}\right)\frac{1}{I_{D2}}-\frac{I_{D1}}{I_{D2}}\frac{1}{V_{\mathrm{DS1}}}
\label{farzan}
\end{equation}
in a first approximation. The RHS of (\ref{farzan}) can be obtained from dc measurements. Once $\partial \overline{Q}_m/\partial V_{\mathrm{DS}}$ is known, the mobility can be extracted by merging (\ref{farzan}) with the drift-diffusion expression for $I_{\mathrm{D2}}$ and eliminating the mobility~\cite{jazaeri}. 

In strong inversion, $\overline{Q}_m$ can be estimated as $\overline{Q}_m=-C_{GG1}(V_{\mathrm{GS1}}-V_{T1})-C_{GG2}(V_{\mathrm{GS2}}-V_{T2})$, with $C_{GG}$ the intrinsic gate capacitance per unit area. This expression is still valid down to deep-cryogenic temperatures since (i) the Maxwell-Boltzmann approximation has been verified down to millikelvin temperatures~\cite{tedpaper}, (ii) in the inversion layer all the dopants are ionized due to field-assisted ionization~\cite{eurosoi,beckers_jeds}, and (iii) interface traps only affect the DC current significantly in the subthreshold region, not in the inversion region~\cite{tedpaper,beckers_jeds}. Therefore, following~\cite{jazaeri} 
\begin{equation}
\frac{\partial \overline{Q}_{m}}{\partial V_{\mathrm{DS}}}=\frac{C_{GG1}+C_{GG2}}{2}. 
\end{equation} 
With (\ref{farzan}), we obtain 
\begin{equation}
\overline{Q}_{m2}\approx \frac{(C_{GG1}+C_{GG2})I_{\mathrm{D2}}}{2\left(\frac{I_{D1}-I_{D2}}{V_{\mathrm{DS1}}-V_{\mathrm{DS2}}}-\frac{I_{D1}}{V_{\mathrm{DS1}}}\right)}
\end{equation}
Thus the mobility is given by 
\begin{equation}
\begin{split}
\mu&\approx -\frac{2L_G}{W(C_{GG1}+C_{GG2})V_{\mathrm{DS2}}}\times\left(\frac{I_{D1}-I_{D2}}{V_{\mathrm{DS1}}-V_{\mathrm{DS2}}}-\frac{I_{D1}}{V_{\mathrm{DS1}}}\right)\\
&=-\frac{2L_G}{W(C_{GG1}+C_{GG2})}\times\frac{1}{V_{\mathrm{DS1}}-V_{\mathrm{DS2}}}\left(\frac{I_{D1}}{V_{\mathrm{DS1}}}-\frac{I_{D2}}{V_{\mathrm{DS2}}}\right)
\end{split}
\label{mob}
\end{equation}
Therefore, for FDSOI the formula remains the same but only the two gate capacitances have to be added. To be very accurate, one would need to consider $C_{GG}(T)$ as a function of temperature, but this would again require CV measurements. However, in the CV characteristics of this technology down to 4.2\,K only a change in the threshold voltage has been observed, and not much change in the shape of $C_{gg}$~\cite{bohus,akturkcapacitance}. Therefore, in strong inversion, we can determine an effective mobility, $\mu_{eff}$, setting $C_{GG1}=\varepsilon_{SiO_2}/\mathrm{EOT}=C_{ox}$ and $C_{GG2}=C_{\mathrm{BOX}}=\varepsilon_{SiO_2}/t_{\mathrm{BOX}. }$. For this reason, in this work we will only investigate mobility-ratios with respect to room temperature, and not the exact values of the mobility. 
To extract the mobility at a constant back-gate voltage, it is allowed to take into account only the capacitance on the front-gate. This gives 
\begin{equation}
\mu_{eff}\approx - \frac{2L_G}{WC_{\mathrm{front\,gate}}}\frac{\partial g_{ds}}{\partial V_{\mathrm{DS}}}, 
\label{gdsfunction}
\end{equation}
which is valid at small $V_{\mathrm{DS}}$ (linear regime) and high $V_{\mathrm{GS}}$ (strong inversion)~\cite{jazaeri}. Expression (\ref{gdsfunction}) is referred to as the $g_{ds}$-function. According to this expression, the mobility is proportional to the curvature of the output characteristics at small $V_{\mathrm{DS}}$ at a given gate voltage and temperature. Note the minus sign, leading to a positive mobility-value since $\partial g_{ds}/\partial V_{\mathrm{DS}}$ is negative. The derivative in (\ref{gdsfunction}) can be calculated from the measurements as the difference in initial slopes using a back-difference method $(g_{ds,1}-g_{ds,0})/(V_{\mathrm{DS1}}-V_{\mathrm{DS0}})$ (at small $V_{\mathrm{DS}}$). The method is versatile since the free-carrier mobility can be obtained either from measured output characteristics using (\ref{gdsfunction}), or from two linear transfer characteristics (\ref{mob}), depending on which low-temperature data is available. Here we extract the mobility from the output characteristics which are available down to 1.4\,K for the long devices, shown in Fig.\ref{figure1}-c and \ref{figure1}-f. 

Figure \ref{figure:mob} plots the ratio of the maximum effective mobility versus the room-temperature mobility for $n$MOS 1\,$\mu m$/1\,$\mu$m (Fig.\,\ref{figure:mob}-a), $p$MOS 1\,$\mu m$/1\,$\mu$m (Fig.\,\ref{figure:mob}-b), and $n$MOS 80\,nm/46\,nm (Fig.\,\ref{figure:mob}-c). A mobility degradation is observed for the long devices, where the temperature with maximum mobility is shifted between $n$MOS ($\approx$10\,K) and $p$MOS ($\approx$77\,K). No mobility degradation is observed on short $n$MOS down to 4.2\,K. 
\section{\label{sec:modeling}Modeling}
In this section, we model the low-temperature measurements (Sec.\,\ref{sec:meas}) using the design-oriented simplified EKV model, focusing on the measurements that do not show any oscillations.

A detailed overview of this model is presented in\cite{enznanoscaleone,enznanoscaletwo}. Its suitability for FDSOI processes has been assessed at room temperature, including body-biasing~\cite{pezzotta}. The model is valid in saturation, expressing the measured drain current in saturation in terms of an inversion coefficient, $IC$, given by $IC\triangleq I_{D,sat} / I_{spec}$, where the specific current, $I_{spec}$, is defined as $I_{spec}\triangleq I_{spec\square}(W/L)$, and the 'specific-current-per-square', $I_{spec\square}$, is a parameter independent of dimensions given by $2n\mu C_{ox}U_T^2$. Many analog figure-of-merits can be expressed in terms of this inversion coefficient, which separates the different regions of inversion as follows: 
\begin{itemize}
	\item $IC < 0.1$: weak inversion
	\item $0.1 < IC < 10$: moderate inversion
	\item $IC > 10$: strong inversion. 
\end{itemize}
The long-channel model is given by the following expression~\cite{enznanoscaleone,enznanoscaletwo}: 

\begin{equation}
v_{p}-v_s=\ln(\sqrt{4IC+1}-1)+\sqrt{4IC+1}-(1+\ln2), 
\label{long}
\end{equation} 
and the short-channel model by
\begin{equation}
\begin{split}
IC&=\frac{4(q_s^2+q_s)}{2+\lambda_c+\sqrt{4(1+\lambda_c)+\lambda_c^2(1+2q_s)^2}},\\
v_p-v_s&=\ln q_s +2q_s,
\end{split}
\label{qs}
\end{equation}
where $v_p\triangleq V_P/U_T$ is the normalized pinch-off voltage, $q_s\triangleq Q_s/Q_{spec}$ the normalized inversion charge at the source (with $Q_{spec}\triangleq -2nU_TC_{ox}$), and $v_s\triangleq V_S/U_T$ the normalized source voltage. The velocity saturation parameter, $\lambda_c=L_{sat}/L$, is the ratio of the channel in full velocity saturation (near the drain) over the total length of the channel. 

Starting from the measured drain current in saturation, the inversion coefficient is evaluated for each $I_{D,sat}$ at a given gate voltage, using a specific model parameter $I_{spec}$. Depending on the length of the channel, we proceed as follows: 

\subsection{Procedure long-channel}
For each $IC$, the normalized pinch-off voltage $v_p$ is obtained from \ref{long}. At a given temperature, the gate voltage follows from $V_g=nU_Tv_p+V_{\mathrm{T0}}$, given specific $n$ and $V_{\mathrm{T0}}$ model parameters. Initial guesses for these model parameters can be obtained from the extracted threshold voltage and slope factors in Fig.\,\ref{fig:char}. 

\subsection{Procedure short-channel}
For each $IC$, the first expression in (\ref{qs}) is numerically solved for $q_s$, given a specific $L_{sat}$. The $v_p$-values are derived from all $q_s$ using the second expression in (\ref{qs}), Similar to the long-channel model, the gate voltage then follows from $V_g=nU_Tv_p+V_{\mathrm{T0}}$, given specific $n$ and $V_{\mathrm{T0}}$ model parameters.\\

Note that the long-channel model uses three model parameters ($n$, $V_{\mathrm{T0}}$, $I_{spec\square}$), while the short-channel model uses four ($n$, $V_{\mathrm{T0}}$, $I_{spec\square}$, $L_{sat}$). By plotting $I_{D,sat}$ versus the obtained $V_g$, the model curves can be validated with the measurements at each temperature, as will be illustrated in the next section. 

\begin{figure}[t]
	\centering
	\includegraphics[scale=0.6]{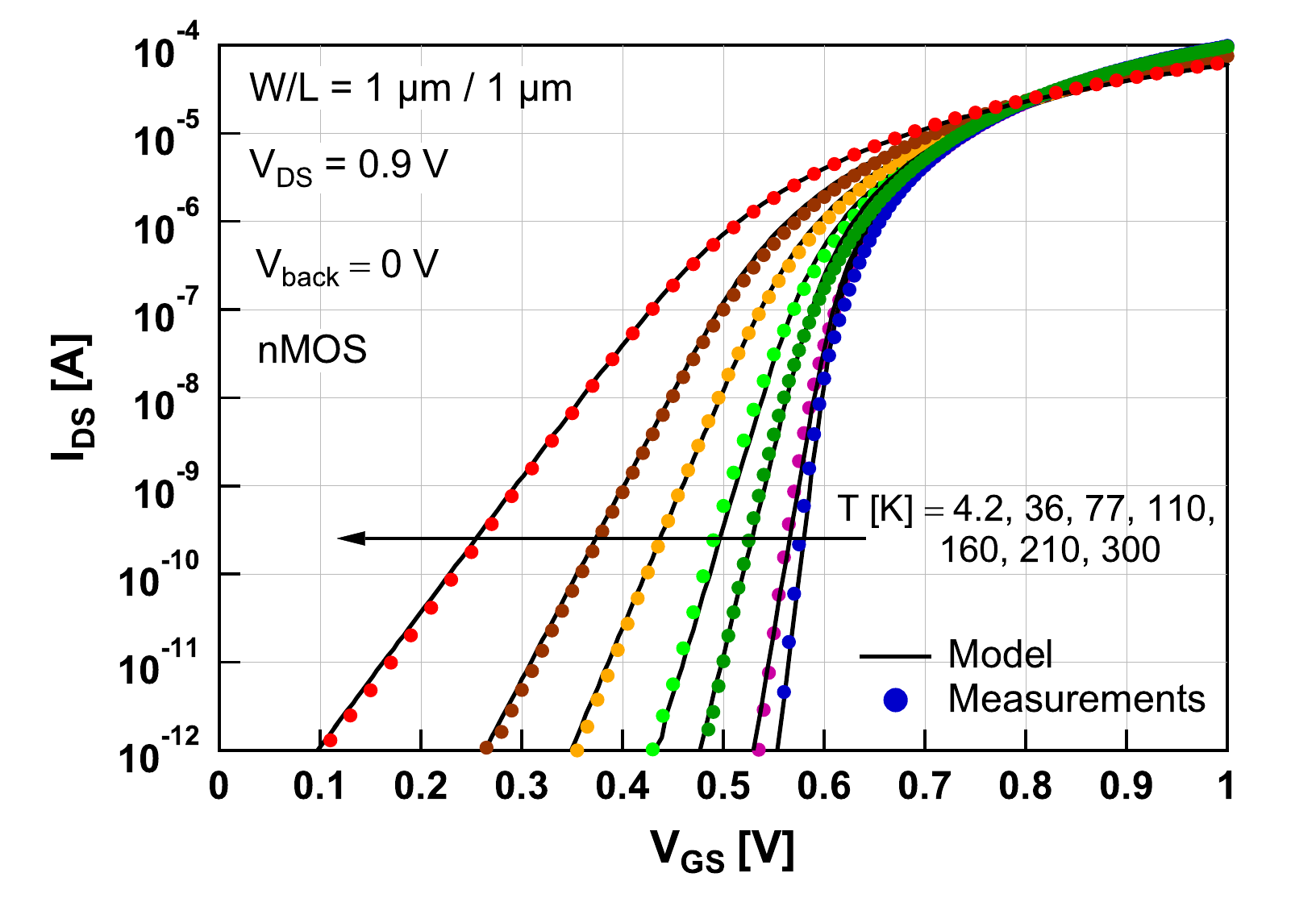}
	\caption{Modeling long FDSOI $n$MOS down to 4.2\,K. Model parameters are given in Table\,\ref{tableNMOS1um1um}.}
	\label{modelingNMOS1um1um}
\end{figure} 
\begin{table}[t]
	\centering
	\begin{tabular}{c c c c}
		\hline
		\textbf{Temperature [K]} & n &$V_{\mathrm{T0}}$ [V] &$I_{spec\square}$ [nA]\\
		\hline
		4.2 & 13 & 0.605 & 55 \\
		36 & 2.1 & 0.6 & 105 \\
		77 & 1.4 & 0.585 & 195\\
		110 & 1.21 & 0.57 & 235 \\
		160 & 1.16 & 0.55 & 395 \\
		210 & 1.1 & 0.525 & 515 \\
	    300 & 1.07 & 0.485 & 835 \\
		\hline
	\end{tabular}
	\caption{\label{tableNMOS1um1um}Model parameters for $n$MOS $W/L=$1\,$\mu m$/1\,$\mu m$ at $V_{back}=$0\,V and increasing temperatures, corresponding to Fig.\,\ref{modelingNMOS1um1um}}
\end{table}

\begin{figure}[t]
	\centering
	\includegraphics[scale=0.6]{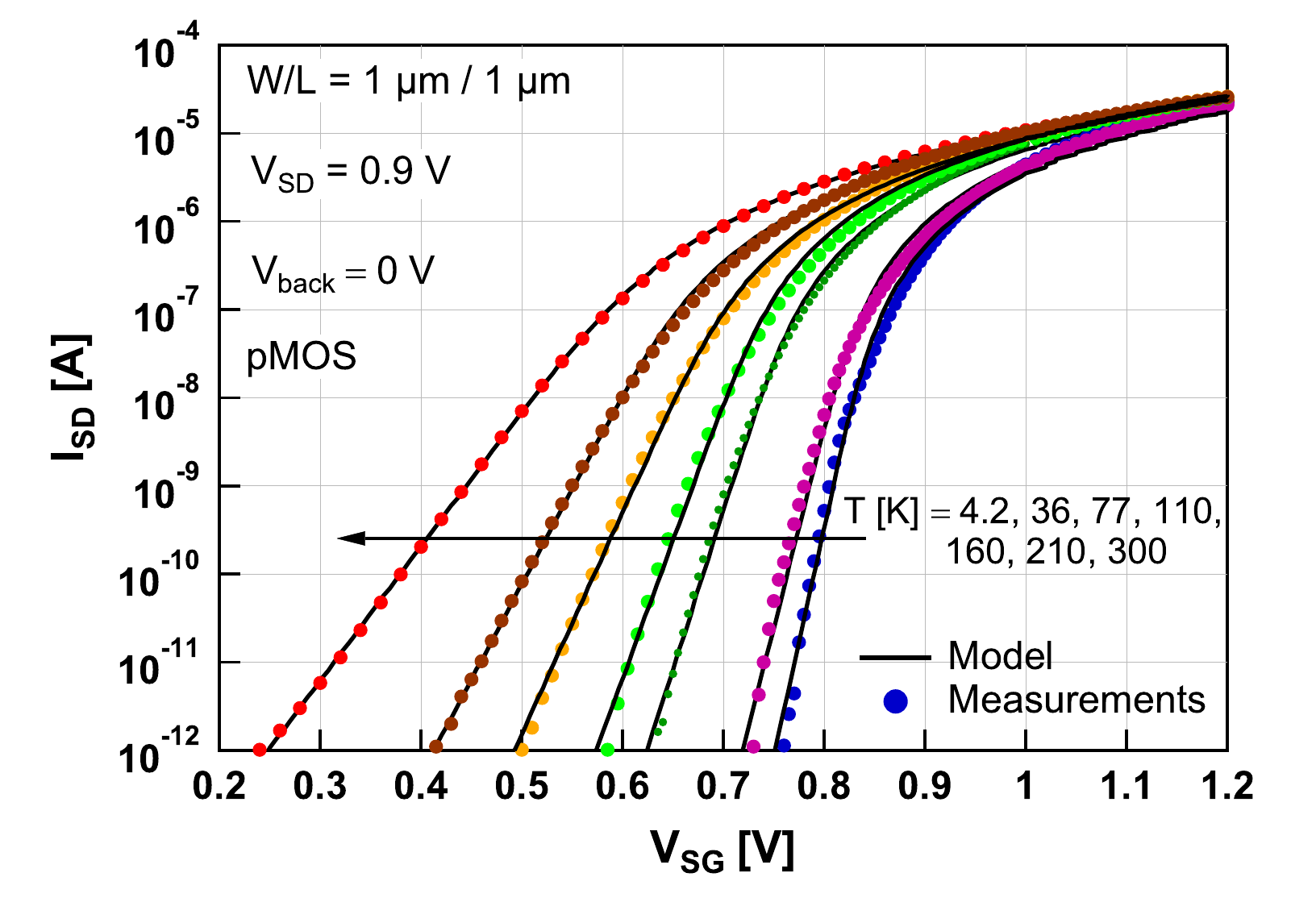}
	\caption{Modeling long FDSOI $p$MOS down to 4.2\,K. Model parameters are given in Table\,\ref{tablepMOS1um1um}.}
	\label{modelingPMOS1um1um}
\end{figure}
\begin{table}[t]
	\centering
	\begin{tabular}{c c c c}
		\hline
		\textbf{Temperature [K]} & n &$V_{\mathrm{T0}}$ [V] &$I_{spec\square}$ [nA]\\
		\hline
		4.2 & 23 & 0.84 & 42 \\
		36 & 3.07 & 0.825 & 65 \\
		77 & 1.82 & 0.76 & 75 \\
		110 & 1.46 & 0.73 & 105 \\
		160 & 1.25 & 0.695 & 125 \\
		210 & 1.11 & 0.65 & 135 \\
		300 & 1.1 & 0.6 & 235 \\
		\hline
	\end{tabular}
	\caption{\label{tablepMOS1um1um}Model parameters for $p$MOS $W/L=$1\,$\mu m$/1\,$\mu m$ at $V_{back}=$0\,V and increasing temperatures, corresponding to Fig.\,\ref{modelingPMOS1um1um}.}
\end{table}

\begin{figure}[t]
	\centering
	\includegraphics[scale=0.6]{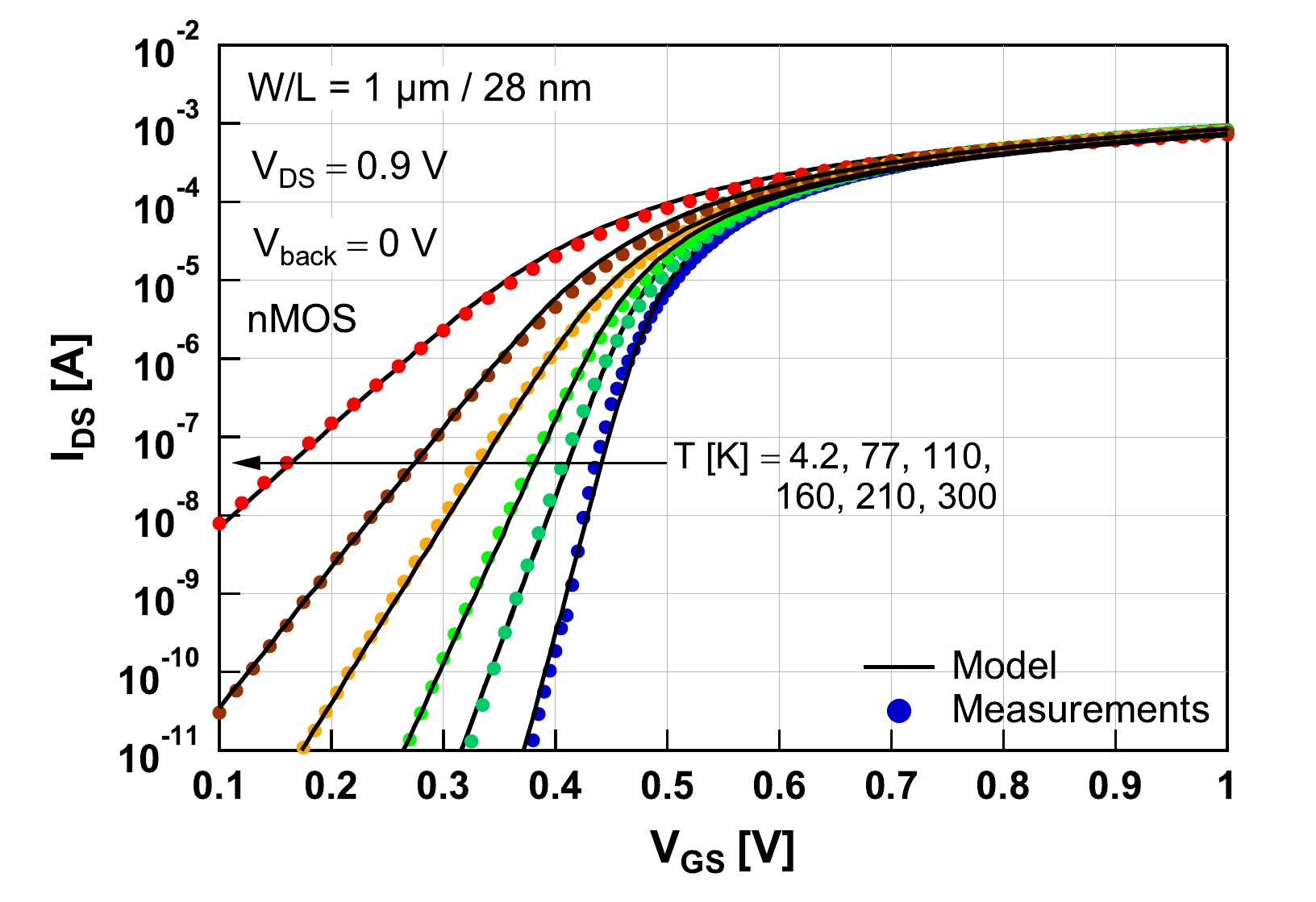}
	\caption{Modeling short 28-nm FDSOI $n$MOS down to 4.2\,K. Model parameters are given in Table\,\ref{tableNMOS1um28nm}.}
	\label{modelingNMOS1um28nm}
\end{figure}

\begin{table}[h]
	\centering
	\begin{tabular}{c c c c c}
		\hline
		\textbf{Temperature [K]} & n &$V_{\mathrm{T0}}$ [V] &$I_{spec\square}$ [nA] & $L_{sat}$ [nm]\\
		\hline
		4.2 & 22 & 0.47 & 75 & 5 \\
		77 & 1.7 & 0.46& 175 & 8\\
		110 & 1.47 & 0.45& 195 & 8.5 \\
		160 & 1.38 & 0.43 & 335 &  9 \\
		210 & 1.34 & 0.41 & 505 & 10 \\
		300 & 1.3 & 0.37 & 835 & 11 \\
		\hline
	\end{tabular}
	\caption{\label{tableNMOS1um28nm}Model parameters for $n$MOS $W/L=$1\,$\mu m$/28\,nm at $V_{back}=$0\,V and increasing temperatures, corresponding to Fig.\,\ref{modelingNMOS1um28nm}.}
\end{table}

\begin{figure}[h]
	\centering
	\includegraphics[scale=0.6]{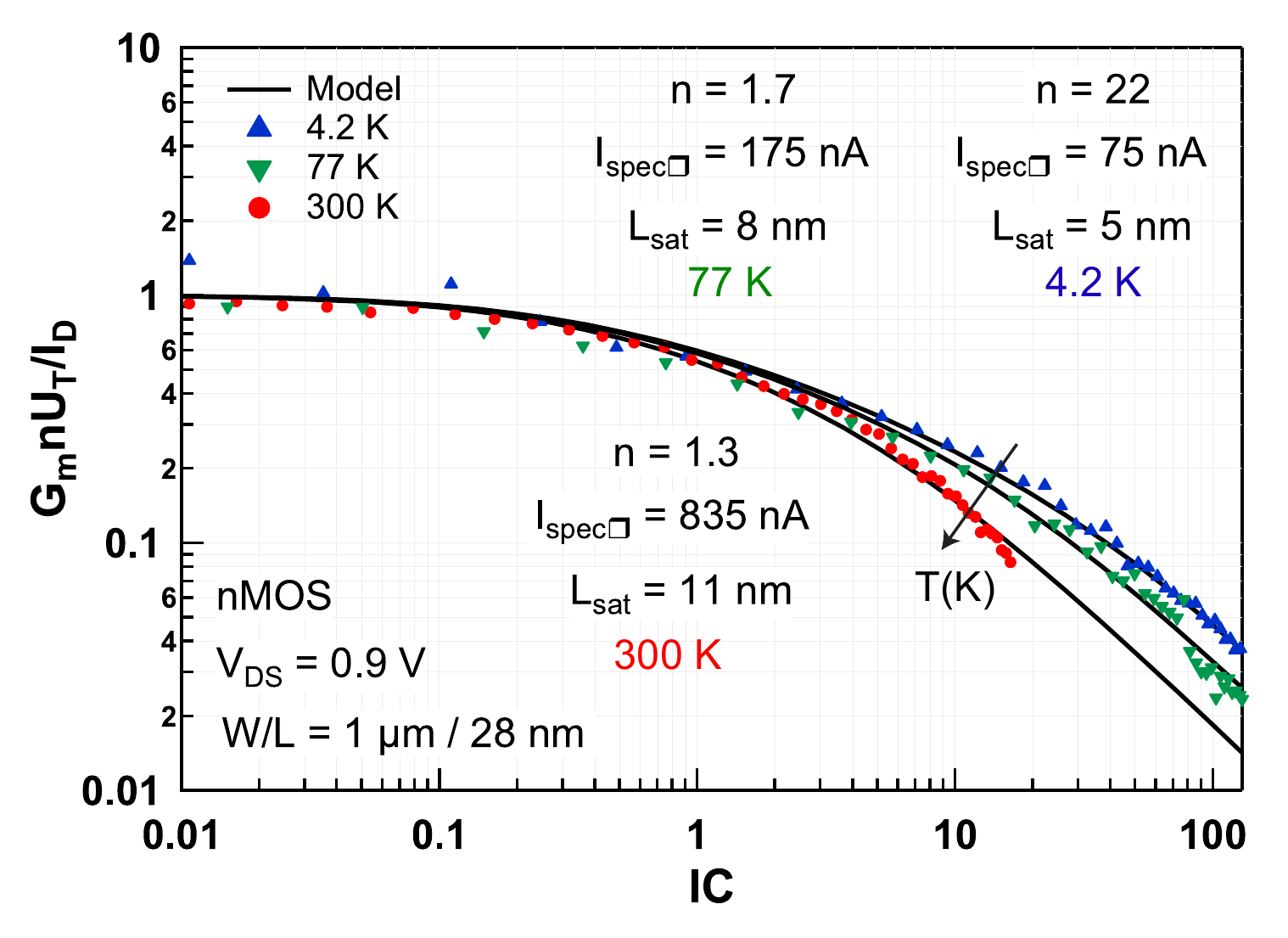}
	\caption{Modeling the normalized transconductance efficiency at 300, 77, and 4.2\,K in a short 28-nm FDSOI $n$MOS in saturation. Model parameters are given in the figure.}
	\label{fig:gmnutid}
\end{figure}
\subsection{Comparison with measurements}
\begin{figure}[t]
	\centering
	\includegraphics[scale=0.6]{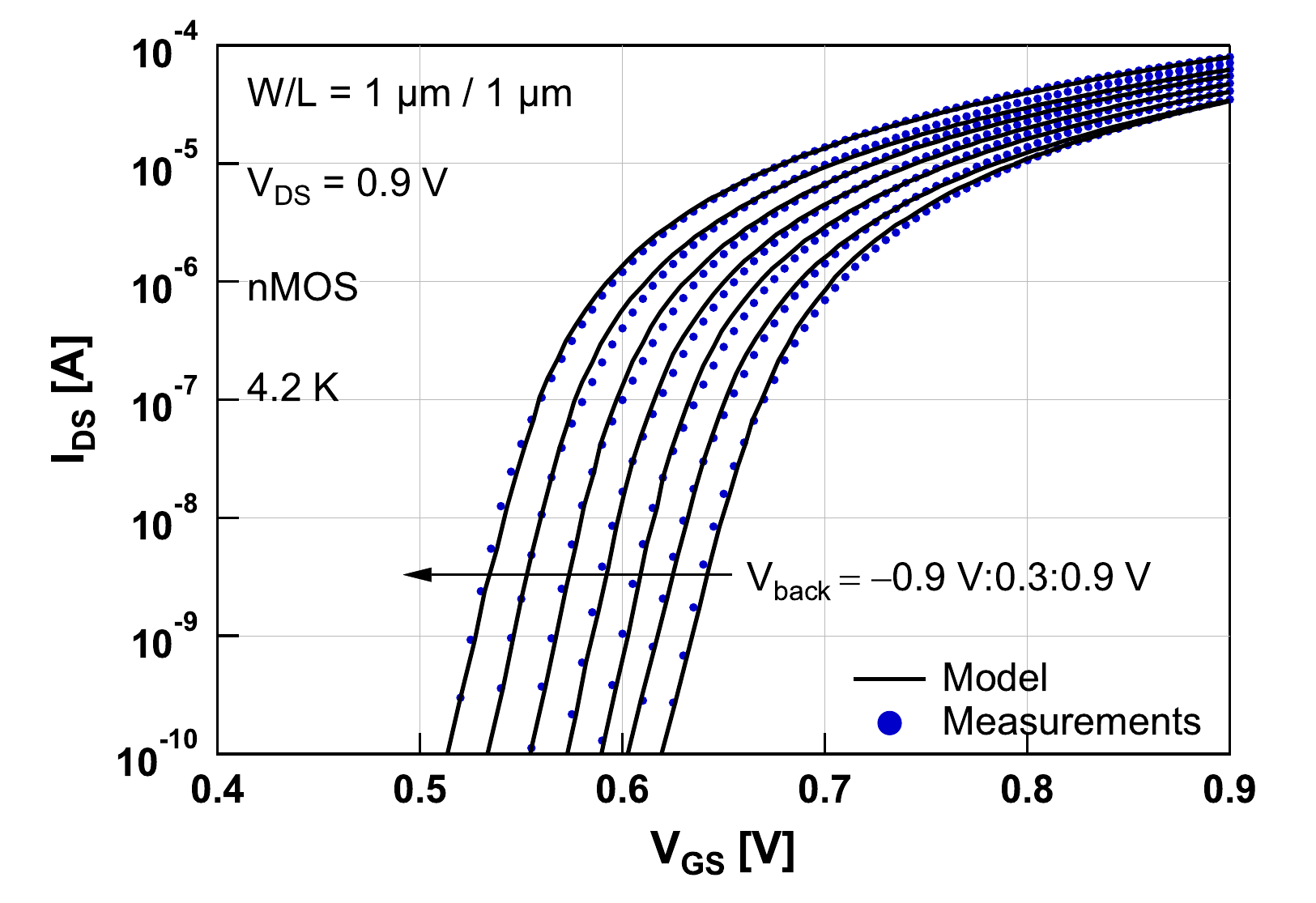}
	\caption{Modeling the body bias effect at 4.2\,K in a long FDSOI $n$MOS. Model parameters are given in Table\,\ref{tableNMOS1um1umVback}.}
	\label{modelingNMOS1um1umVback}
\end{figure}

\begin{figure}[!ht]
	\centering
	\includegraphics[scale=0.6]{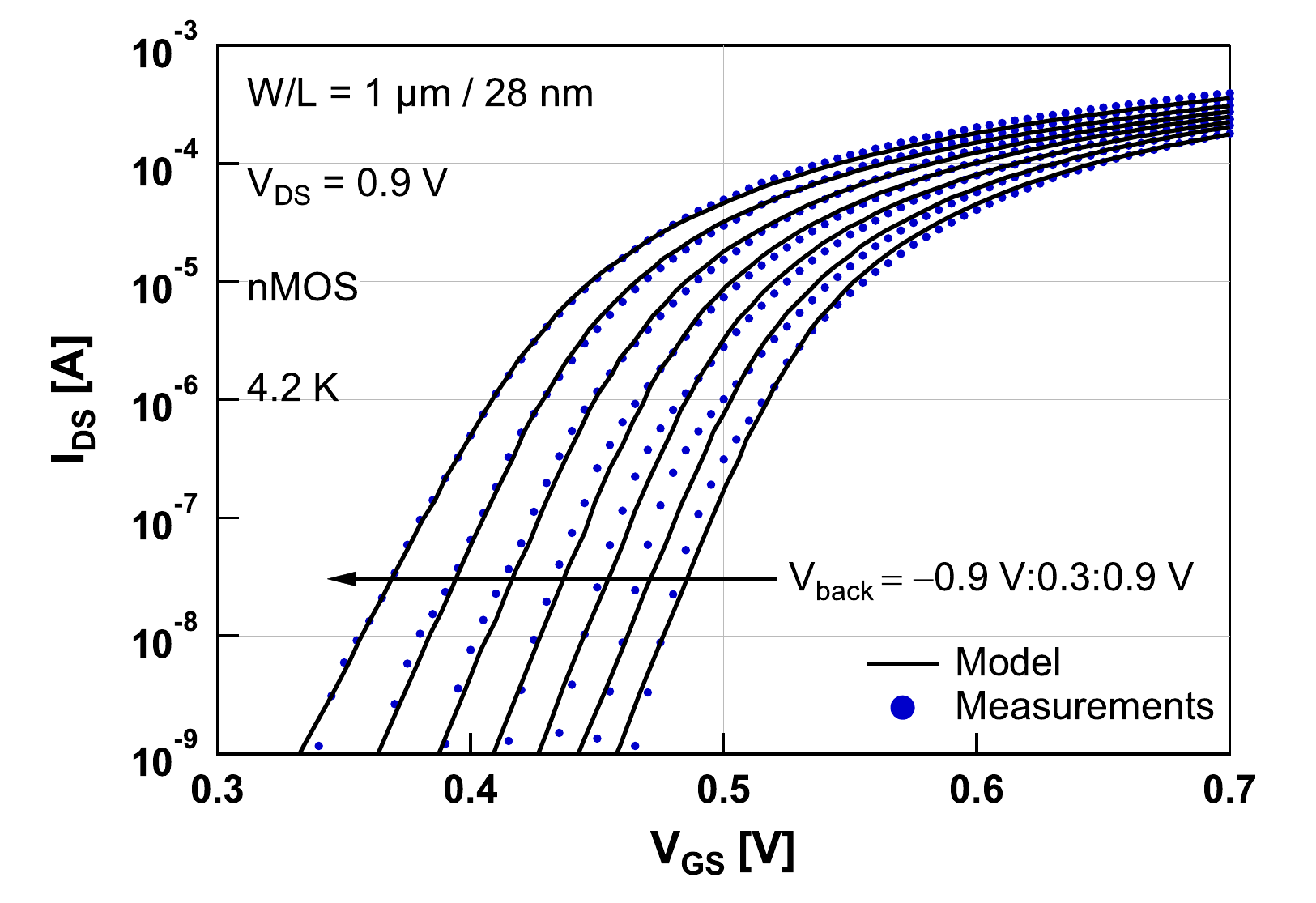}
	\caption{Modeling the body bias effect at 4.2\,K in a short 28-nm FDSOI $n$MOS. Model parameters are given in Table\,\ref{tableNMOS1um28nmVback}.}
	\label{modelingNMOS1um28nmVback}
\end{figure}

Using the model over a wide temperature range (from 300 down to 4.2\,K), the transfer characteristics, back-gate sensitivity, and transconductance efficiency can be accurately modeled in long and short FDSOI devices. Figures \ref{modelingNMOS1um1um} to \ref{modelingNMOS1um28nm} show the modeled transfer characteristics in saturation at all considered temperatures down to 4.2\,K at zero back-gate voltage for long $n$MOS (Fig.\,\ref{modelingNMOS1um1um}), long $p$MOS (Fig.\,\ref{modelingPMOS1um1um}), and short $n$MOS (Fig.\,\ref{modelingNMOS1um28nm}). The model parameters are shown in the tables below the figures. The strong increase in the $n$ model-parameter at deep-cryogenic temperatures corresponds to the interface-trapping process, as explained in Sec.\,\ref{sec:char}. The $V_{\mathrm{T0}}$ model-parameter captures the change in the threshold voltage due to Fermi-Dirac scaling and incomplete ionization, increasing in the order of $0.1$\,V. Note that the used values for $n$ and $V_{\mathrm{T0}}$ correspond to the extracted values in Fig.\,\ref{fig:char}-b and \ref{fig:char}-c. The $I_{spec\square}$ model-parameter decreases over one order of magnitude from 300 down to 4.2\,K. For the short device, the $L_{sat}$-parameter decreases from 11 to 5\,nm due to a reduction in the phonon scattering, leading to a shorter part of the channel near the drain in velocity saturation. The lower impact of velocity saturation at lower temperatures becomes clear also by plotting the normalized transconductance efficiency, $G_mnU_T/I_D$, versus the inversion coefficient at 300, 77, and 4.2\,K, shown in Fig.\,\ref{fig:gmnutid}. Using the same parameters for $n$, $I_{spec\square}$, and $L_{sat}$ as in Fig.\,\ref{modelingNMOS1um28nm}, good agreement is obtained between the modeled and measured transconductance efficiency at 300, 77, and 4.2\,K. Fig.\,\ref{fig:gmnutid} verifies that the $G_m/I_D$ design-methodology\cite{silveira1996g} remains valid for a 28\,nm FDSOI technology down to 4.2\,K, extending therefore its universality to advanced bulk and FDSOI CMOS operating at extremely-low temperatures. Furthermore, as illustrated in Figures \,\ref{modelingNMOS1um1umVback} and \ref{modelingNMOS1um28nmVback}, changing the $V_{\mathrm{T0}}$ model parameter allows to capture the effect of the back-gate at 4.2\,K for long (\ref{modelingNMOS1um1umVback}) and short devices. The $n$ model parameter tends to increase with increasing absolute values of the back-gate voltage in both long and short devices, accounting for a change in $SS$ induced by the back gate. The $L_{sat}$ model parameter maintains the same value (5\,nm at 4.2\,K) for different back-gate voltages, showing that the velocity saturation is not influenced by the back gate.   

\section{Conclusion}
A 28-nm Fully-Depleted SOI CMOS process is characterized and modeled from room temperature down to liquid-helium temperature (4.2\,K). Output characteristics and free-carrier mobilities are presented down to 1.4\,K. The design-oriented simplified EKV model can accurately predict the impact of the temperature reduction on the transfer characteristics, back-gate sensitivity, and transconductance efficiency of 28-nm devices using four parameters: the slope factor $n$, threshold voltage $V_{\mathrm{T0}}$, specific current $I_{spec}$, and saturation length $L_{sat}$. A new method is proposed to extract the free-carrier mobility-trends versus temperature in SOI technology from dc measurements. This method does not require CV measurements and can hence be used to extract the mobility-trend also on short-narrow advanced CMOS devices where parasitic capacitances can dominate. Using this method, a degradation in the free-carrier mobility is observed at cryogenic temperatures in long $n$MOS and $p$MOS, and an increase in a short 46-nm $n$MOS.  

\begin{table}[t]
	\centering
	\begin{tabular}{c c c c}
		\hline
		\textbf{Back-gate Voltage [V]}  & n &$V_{\mathrm{T0}}$ [V] &$I_{spec\square}$ [nA]\\
		\hline
		-0.9 & 18 & 0.665 & 115 \\
		-0.6 & 17.2 & 0.645 & 90 \\
		-0.3 & 14.9 & 0.625 & 68\\
		0 & 14.9 & 0.608 & 70 \\
		0.3 & 15 & 0.59 & 73 \\
		0.6 & 15.2 & 0.57 & 75 \\
		0.9 & 16 & 0.55 & 105 \\
		\hline
	\end{tabular}
	\caption{\label{tableNMOS1um1umVback}Model parameters for $n$MOS $W/L=$1\,$\mu m$/1\,$\mu m$ at 4.2\,K and ramping $V_{back}$, corresponding to Fig.\,\ref{modelingNMOS1um1umVback}.}
\end{table}

\begin{table}[t]
	\centering
	\begin{tabular}{c c c c c}
		\hline
		\textbf{Back-gate Voltage [V]} & n &$V_{\mathrm{T0}}$ [V] &$I_{spec\square}$ [nA] & $L_{sat}$ [nm]\\
		\hline
		-0.9 & 22.8 & 0.522 & 77 & 5\\
		-0.6 & 22.7 & 0.506 & 76 & 5\\
		-0.3 & 22.6 & 0.49 & 75 & 5\\
		0 & 22.7 & 0.473 & 70 & 5\\
		0.3 & 23.9 & 0.455 & 80 & 5\\
		0.6 & 25 & 0.435 & 85 & 5\\
		0.9 & 29.3 & 0.42 & 115 & 5\\
		\hline
	\end{tabular}
	\caption{\label{tableNMOS1um28nmVback}Model parameters for $n$MOS $W/L=$1\,$\mu m$/28\,nm at 4.2\,K and ramping $V_{back}$, corresponding to Fig.\,\ref{modelingNMOS1um28nmVback}.}
\end{table}

\section*{References}

\bibliography{mybibfile}

\end{document}